\documentclass{PoS}
\usepackage{epsfig}
\usepackage{verbatim, amsmath,amssymb}

\newcommand{\eq}[1]{Eq.~(\ref{#1})}
\newcommand{\fig}[1]{Fig.~{\ref{#1}}}

\newcommand{\be}{\begin{equation}}
\newcommand{\ee}{\end{equation}}
\newcommand{\bea}{\begin{eqnarray}}
\newcommand{\eea}{\end{eqnarray}}
\newcommand{\ben}{\begin{eqnarray*}}
\newcommand{\een}{\end{eqnarray*}}

\newcommand{\DS}{Dyson-Schwinger }
\newcommand{\BS}{Bethe-Salpeter }
\newcommand{\ST}{Slavnov-Taylor }
\newcommand{\YM}{Yang-Mills }

\newcommand{\w}{\omega}
\newcommand{\e}{\varepsilon}
\newcommand{\al}{\alpha}
\newcommand{\ba}{\beta}
\newcommand{\ga}{\gamma}
\newcommand{\G}{\Gamma}
\newcommand{\de}{\delta}

\newcommand{\si}{\sigma}

\newcommand{\la}{\lambda}

\newcommand{\ka}{\kappa}

\newcommand{\cs}{{\cal S}}
\renewcommand{\div}{\vec{\nabla}}

\newcommand{\s}[2]{{#1}\!\cdot\!{#2}}
\newcommand{\ov}[1]{\overline{#1}}
\newcommand{\dk}[1]
{\,\,\,\raisebox{-0.4ex}{\large $\bar{}$}\!\!d\,{#1}\,}

\newcommand{\ev}[1]{<\!\!{#1}\!\!>}

\title{Nonperturbative study of the four-point heavy quark Green's
functions in Coulomb gauge}

\ShortTitle{Nonperturbative study of the four-point heavy quark 
Green's functions}

\author{\speaker{Carina Popovici}\\
Departamento de F\'{\i}sica,Universidade de Coimbra,
 3004-516 Coimbra, Portugal\\
        E-mail: \email{popovici@teor.fis.uc.pt}}
\author{Peter Watson, Hugo Reinhardt\\
Institut f\"ur Theoretische Physik, Universit\"at
T\"ubingen, Auf der Morgenstelle 14, D-72076 T\"ubingen, Germany
}

\abstract{
The heavy quark sector of Coulomb gauge QCD is investigated, by making
a heavy quark mass expansion of the QCD action and restricting to the
leading order. With the truncation of the Yang-Mills sector to include
only dressed two-point functions, we study the Dyson-Schwinger
equations for the four-point quark Green's functions (proper and
amputated). In this limit, we provide an exact solution for the
four-point quark Green's functions and show that the corresponding
poles relate to the bound state energy of the heavy quark system. 
Moreover, a natural separation between the physical and unphysical
poles in the Green's functions emerges.}

\FullConference{International Workshop on QCD Green's Functions,
Confinement and Phenomenology\\5-9 September 2011\\ Trento, Italy}

\begin{document}

\section{Introduction}

Coulomb gauge is a suitable choice for investigating the confinement
phenomenon, as in this gauge the Gribov-Zwanziger scenario becomes
distinctly prominent: the temporal component of the gluon propagator
provides the long range confining potential, whereas the spatial
propagator is infrared suppressed \cite{Gribov:1977wm}. On the other
hand, traditional studies of the heavy quark sector of quantum
chromodynamics (QCD) mainly use phenomenologically motivated
potentials in the place of the \YM sector. In this context, it is
appropriate to investigate the relationship between the
nonperturbative scale associated with confinement and the \YM sector
of the theory.

This talk reviews results obtained in the heavy quark limit of Coulomb
gauge QCD \cite{Popovici:2010mb, arXiv:1103.4786}.  After making an
expansion in the heavy quark mass and restricting to the leading
order, we consider the heavy quark propagator and the homogeneous \BS
equation for quark-antiquark systems. With the further truncation to
exclude pure \YM vertices (but retaining nonperturbative dressed
propagators), we show that the rainbow-ladder approximation is exact
in this case and we establish a connection between the temporal gluon
propagator and the external physical scale (the string tension), at
least within the leading order truncation. In the second part of the
talk, we investigate the (full nonperturbative) four-point
quark-antiquark Green's functions. We present exact, analytic
solutions, and show that the physical poles of the Green's function
explicitly separate from the possible unphysical ones. Moreover, we
find that the physical poles of the \BS equation are contained within
the singularities of the Green's function. These results will
hopefully be useful in the future investigations of phenomenological
models for mesons and baryons (see, for example,
Ref.~\cite{Blank:2010bp} for a numerical analysis of the inhomogeneous
\BS equation).

\section {Quark propagator in the heavy mass limit}

Let us start by considering the explicit quark contribution to the full QCD
generating functional:
\bea 
Z[\ov{\chi},\chi]&=&\int{\cal D}\Phi\exp{\left\{\imath \int d^4x
\ov{q}_\al(x)\left[\imath\ga^0D_0
+\imath\s{\vec{\ga}}{\vec{D}}-m\right]_{\al\ba}q_\ba(x) \right\}}
\nonumber\\&&
\times \exp{\left\{\imath\int
d^4x\left[\ov{\chi}_\al(x)q_\al(x)+
\ov{q}_\al(x)\chi_\al(x)\right]+\imath \cs_{YM}\right\}}.
\label{eq:genfunc} 
\eea
In the above, ${\cal D}\Phi$ denotes the integration over all fields
present, $q_\al$ is the quark field, $\bar q_{\al}$ the conjugate
antiquark field, and $\bar\chi_{\al},\chi_{\al}$ the corresponding
sources. The common index $\al,\ba\dots$ denotes the color, spin and
flavor indices.  The Dirac $\gamma$ matrices satisfy
$\left\{\ga^{\mu},\ga^{\nu}\right\}=2g^{\mu\nu}$, with the metric
$g^{\mu\nu}=\textrm {diag} (1,-\vec 1)$. The structure constants of
the $SU(N_c)$ group are denoted with $f^{abc}$, and the Hermitian
generators satisfy $[T^a,T^b]=\imath f^{abc} T^{c}$ and are normalized
via $\textrm {Tr} (T^a,T^b)=\de^{ab}/2$.  $\cs_{YM}$ represents the
\YM part of the action and 
\bea
D_0=\partial_{0}-\imath gT^a\si^a(x),
\,\,\,\,\,\, \vec{D}=\div+\imath gT^a\vec{A}^a(x), 
\eea
are the temporal and spatial components of the covariant derivative in
the fundamental color representation ($\vec{A}$ and $\si$ refer to the
spatial and temporal components of the gluon field, respectively).

 We now decompose the  full quark field according to the heavy
quark transformation 
\be
q_\al(x)=e^{-\imath mx_0}\left[h(x)+H(x)\right]_\al,\;
h_\al(x)=e^{\imath mx_0}\left[\frac{{\mathbf 1}
+\ga^0}{2}q(x)\right]_\al,\;
H_\al(x)=e^{\imath mx_0}\left[\frac{{\mathbf 1}
-\ga^0}{2}q(x)\right]_\al
\label{eq:qdecomp}
\ee
(similarly for the antiquark field), where the two components $h$ and
$H$ are introduced with the help of the spinor projectors $(1\pm
\ga^0)/2$. This is a particular case of the heavy quark transform,
adopted from the Heavy Quark Effective Theory [HQET]
\cite{Neubert:1993mb}, which turns out to be useful in Coulomb gauge.

After inserting the quark fields, decomposed according to
\eq{eq:qdecomp}, into the generating functional \eq{eq:genfunc}, we
integrate out the $H$-fields and make an expansion in the inverse of
the heavy quark mass (in the following, we will adopt the standard
terminology and denote it simply ``mass expansion'').  At leading
order, the generating functional reduces to:
\bea
Z[\ov{\chi},\chi]&=&\int{\cal D}\Phi\exp{\left\{\imath
\int d^4x\ov{h}_\al(x)
\left[\imath\partial_{0x}+gT^a\si^a(x)\right]_{\al\ba}
h_\ba(x)\right\}}
\nonumber\\&&\times
\exp{\left\{\imath
\int d^4x\left[e^{-\imath mx_0}\ov{\chi}_\al(x) h_\al(x)
+e^{\imath mx_0}\ov{h}_\al(x)\chi_\al(x)\right]+
\imath{\cal S}_{YM}\right\}}+{\cal O}\left(1/m\right),
\label{eq:genfunc4}
\eea
where we have replaced the covariant derivative $D_0$ with its
explicit expression.  In the above expression, we notice that as a
result of the heavy quark transformation, at leading order in the mass
expansion the quark interacts only with the temporal gluon, whereas
the spatial component is suppressed.  Also, note the absence of the
Dirac $\ga$ structure, as a result of the multiplication with the
projectors $({\mathbf 1}\pm\ga^0)/2$ (physically, this implies that
the spin degree of freedom decouples from the system). A further
important point is that in \eq{eq:genfunc4} we have kept the full
quark and antiquark sources, as opposed to HQET, where the sources
corresponding to the large components $h$ are used. This means that at
leading order in the mass expansion we are allowed to use the full
apparatus of the functional formalism, and hence derive the full \DS
equations in Coulomb gauge QCD, while replacing the corresponding
propagators and vertices by their leading order expressions.

In Coulomb gauge QCD (without the mass expansion), the quark gap
equation for the proper two-point function is given by (see
Ref.~\cite{Popovici:2008ty} for a complete derivation and notation):
\bea
\G_{\ov{q}q\al\ga}(k)&=\G_{\ov{q}q\al\ga}^{(0)}(k)&+\!\int\!\dk{\w}
\left\{\G_{\ov{q}q\si\al\ba}^{(0)a}(k,-\w,\w-k)W_{\ov{q}q\ba\ka}
(\w)\G_{\ov{q}q\si\ka\ga}^{b}(\w,-k,k-\w)W_{\si\si}^{ab}(k-\w) \right.
\nonumber\\&&\left.
\!+\G_{\ov{q}qA\al\ba i}^{(0)a}(k,-\w,\w-k)W_{\ov{q}q\ba\ka}(\w)
\G_{\ov{q}qA\ka\ga j}^{b}(\w,-k,k-\w)W_{AAij}^{ab}(k-\w) \right\}\!,
\label{eq:gap}
\eea
where $\dk{\w}=d\w/(2\pi)^4$.  In order to solve this equation, we
first need to specify the tree-level quark proper two-point function
and the components of the tree-level quark-gluon vertex. They are
derived directly from the generating functional \eq{eq:genfunc4} and
are given by:
\bea
&&\G_{\ov{q}q\al \ba}^{(0)}(k)=\imath\de_{\al\ba}\left[k_0-m\right]
+{\cal O}\left(1/m\right),
\label{eq:gaptree}\\
&&\G_{\ov{q}q\si\al\ba}^{(0)a}(k_1,k_2,k_3)=
\left[gT^a\right]_{\al\ba}+{\cal O}\left(1/m\right).
\label{eq:feyn0}
\eea
Note that the spatial component of the quark-gluon vertex is of order
${\cal O} (1/m)$, as shall be explained shortly below. Further, the
nonperturbative temporal gluon propagator entering \eq{eq:gap} has the
form \cite{Watson:2007vc}:
\be
W_{\si\si}^{ab}(k)=\de^{ab} W_{\si\si}(\vec k)
=\de^{ab}\frac{\imath}{\vec{k}^2}D_{\si\si}(\vec{k}^2).
\label{eq:Wsisi}
\ee
Following lattice results, which signal that that the gluon dressing
function is largely independent of energy, we assume that $D_{\si\si}$
is a function of the three-momentum. Also, lattice investigations
indicate that $D_{\si\si}$ is infrared divergent and behaves like
$1/\vec{k}^2$ for vanishing $\vec{k}^2$ \cite{Quandt:2008zj} (however,
we will need the explicit form of this function only at the end of the
calculation).

Finally, the last input is provided by the \ST identity, which
furnishes a relation between the two- and three-point functions of the
theory.  This is derived from the invariance of the QCD action under a
time-dependent Gauss-BRST transform \cite{Popovici:2010mb}, and in
Coulomb gauge reads ($k_1+k_2+k_3=0$):
\bea
k_3^0\G_{\ov{q}q\si\al\ba}^{d}(k_1,k_2,k_3)&=&
\imath\frac{k_{3i}}
{\vec{k}_3^2}\G_{\ov{q}qA\al\ba i}^{a}(k_1,k_2,k_3)
\G_{\ov{c}c}^{ad}(-k_3)\nonumber\\&&
+\G_{\ov{q}q\al\de}(k_1)\left[\tilde{\G}_{\ov{q};\ov{c}cq}^{d}
(k_1+q_0,k_3-q_0;k_2)+\imath gT^d\right]_{\de\ba}
\nonumber\\&&
+\left[\tilde{\G}_{q;\ov{c}c\ov{q}}^{d}(k_2+q_0,k_3-q_0;k_1)-
\imath gT^d\right]_{\al\de}\G_{\ov{q}q\de\ba}(-k_2).
\label{eq:stid}
\eea
In the above, $q_0$ is an arbitrary energy injection scale (arising
from the noncovariance of Coulomb gauge \cite{Watson:2008fb}),
$\G_{\ov{c}c}$ is the ghost proper two-point function, and
$\tilde{\G}_{\ov{q};\ov{c}cq}$ and $\tilde{\G}_{q;\ov{c}c\ov{q}}$ are
ghost-quark kernels associated with the time-dependent Gauss-BRST
transform.  Since in the generating functional, \eq{eq:genfunc4}, the
tree-level spatial quark-gluon vertex appears at ${\cal O}(1/m)$, by
making the further truncation to neglect the pure \YM vertices, it
follows from the \DS equation for the spatial quark-gluon vertex that
the fully dressed $\G_{\bar qqA}\sim {\cal O}(1/m)$ (and hence it is
neglected).  Further, the ghost-gluon vertices involve pure \YM
vertices and hence are also truncated out.  Thus, in our truncation
scheme and at leading order in the mass expansion, the \ST identity
takes the simple form:
\be
k_3^0\G_{\ov{q}q\si\al\ba}^{d}(k_1,k_2,k_3)=
\G_{\ov{q}q\al\de}(k_1)\left[\imath gT^d\right]_{\de\ba}-
\left[\imath gT^d\right]_{\al\de}\G_{\ov{q}q\de\ba}(-k_2)
+{\cal O}\left(1/m\right).
\label{eq:stidlimit}
\ee

Collecting the above results, we find the following expression for the
heavy quark propagator, as a solution of \eq{eq:gap}, combined with
\eq{eq:stidlimit}:
\bea
W_{\ov{q}q\al\ba}(k)=\frac{-\imath\de_{\al\ba}}{\left[k_0-m-
{\cal I}_r+\imath\e\right]}+{\cal O}\left(1/m\right),
\label{eq:quarkpropnonpert}
\eea
with the constant (implicitly regularized, as indicated 
  by the index ``$r$''):
\be
 {\cal I}_r =\frac{1}{2}g^2C_F
\int_r\frac{\dk{\vec{\w}}D_{\si\si}(\vec{\w})}{\vec{\w}^2}
+{\cal O}\left(1/m\right),
\label{eq:iregularized}
\ee
where $\dk{\vec{\w}}=d^3\vec{\w}/(2\pi)^3$ and $C_F=(N_c^2-1)/2N_c$.
The gap equation, \eq{eq:gap}, has been solved under the assumption
that the temporal integral has been performed first, with the spatial
integral regularized and finite. The solution, 
\eq{eq:quarkpropnonpert}, is then inserted into the \ST identity,
\eq{eq:stidlimit}, and we find that the temporal quark-gluon vertex
remains nonperturbatively bare:
\be
\G_{\ov{q}q\si\al\ba}^{a}(k_1,k_2,k_3)=
\left[gT^a\right]_{\al\ba}+{\cal O}\left(1/m\right).
\label{eq:feyn}
\ee

Note that the quark propagator, \eq{eq:quarkpropnonpert}, possesses a
single pole in the complex $k_0$ plane, in contrast with the standard
QCD propagator, which has a double covariant pole. Hence, we need to
derive the quark and antiquark propagators separately, with the
corresponding Feynman prescriptions. Examining the closed quark loops
(virtual quark-antiquark pairs connected by a primitive vertex), we
find that they vanish due to the energy integration over two quark
propagators with the same Feynman prescription, and this implies that
the theory is quenched in the heavy mass limit:
\be
\int\frac{dk_0}{\left[k_0-m- {\cal I}_r+\imath\e\right]
\left[k_0+p_0-m- {\cal I}_r+\imath\e\right]}=0.
\label{eq:tempint}
\ee
Further, note that the propagator \eq{eq:quarkpropnonpert} is diagonal
in the outer product of the fundamental color, flavor and spinor
spaces, and this exhibits the decoupling of the spin from the heavy
quark system.  Finally, it is important to emphasize that the position
of the pole has no physical meaning, since not the propagator itself,
but the bound state of a quark and an antiquark is physical. The fact
that the poles in the quark propagator are shifted to infinity once
the regularization is removed simply means that an infinite energy is
needed to create a single quark from the vacuum. If a hadronic state
is considered, only the relative energy (derived from the homogeneous
\BS equation) is required to describe the system, and in this case the
singularities appearing in \eq{eq:quarkpropnonpert} cancel. Similar
types of cancellation appear in the solutions of the four-point
quark-antiquark Green's function (see also the discussion from section
\ref{sec:4pgf}).

For the antiquark propagator we obtain:
\bea
W_{q\ov{q} \al\ba}(k)=\frac{-\imath \de_{\al\ba}} {\left[k_0+m-{\cal
I}_r+\imath\e\right]} +{\cal O}\left(1/m\right),
\label{eq:antiquarkpropnonpert}
\eea
and the corresponding temporal antiquark-gluon vertex is given by:
\be
\G_{q\ov{q}\si \al\ba}^{a}(k_1,k_2,k_3)
=-\left[gT^a\right]_{\ba\al}+{\cal O}\left(1/m\right).
\label{eq:antiqqsinp}
\ee
In the above, notice the Feynman prescription of the propagator, as
well as the sign of the loop correction, ${\cal I}_r$. As shall be
discussed in the next section, this apparently minor modification will
play an important role in the interpretation of the solutions of the
\BS equation for quark-antiquark states as bound state/confining
solutions.

\section{Homogeneous \BS equation}

\begin{figure}[t]
\centering\includegraphics[width=0.42\linewidth]{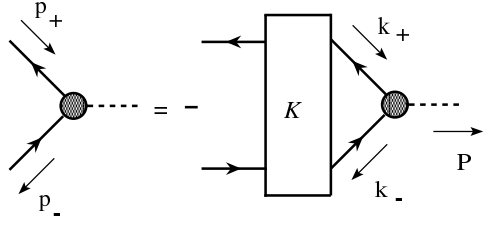}
\caption{\label{fig:fbse}Homogeneous \BS equation for 
quark-antiquark bound states.  Internal propagators are fully 
dressed and solid lines represent the quark propagator.  The box 
represents the \BS kernel $K$ and filled blobs represent the \BS 
vertex function $\G$ with the (external) bound state leg given by 
a dashed line.}
\end{figure}

Let us now consider the full homogeneous \BS equation for
quark-antiquark bound states \cite{Popovici:2010mb} (see also
\fig{fig:fbse}):
\be
\G(p;P)_{\al\ba}=-\int\dk{k}
K_{\al\ba;\de\ga}(p,k;P)\left[W_{\ov{q}
q}(k_+)\G(k;P)W_{\ov{q}q}(k_-)\right]_{\ga\de}.
\label{eq:bseq}
\ee
In the above, the momenta of the quarks are given by $p_+=p+\xi P$,
$p_-=p-(1-\xi) P$ (similarly for $k_{\pm}$), and $\xi$ is the momentum
sharing fraction (note that the solutions here and in the next section
turn out to be independent of $\xi$, just as in the covariant case
\cite{hep-ph/0202053}).  $P$ represents the 4-momentum of the bound
state (assuming that a solution exists), $\G$ is the \BS vertex
function for the particular bound state and $K$ is the \BS kernel,
which still needs to be specified.

It is well-known that the \BS kernel $K$ and the quark self-energy are
related via the axialvector Ward-Takahashi identity
\cite{Adler:1984ri}. Under truncation, it was seen in the last section
that the heavy quark self-energy reduces to the rainbow truncated
form; the corresponding truncation for the kernel is the ladder
approximation. In \cite{Popovici:2010mb}, this has been explicitly
derived.  The \BS kernel in the heavy quark-antiquark system under the
truncation considered here is thus:
\be
K_{\al\ba;\de\ga}(p,k)= 
\G_{\bar qq\si\al\ga}^{a}(p_+,-k_+,k-p)W_{\si\si}^{ab}(\vec p-\vec k)
\G_{q\bar q\si\ba\de}^{bT} (-p_-,k_-,p-k).
\label{eq:kernel}
\ee

We now insert the nonperturbative results for the propagators and
vertices, Eqs.~(\ref{eq:quarkpropnonpert},\ref{eq:feyn},
\ref{eq:antiquarkpropnonpert},\ref{eq:antiqqsinp}) and the expression
\eq{eq:Wsisi} for the temporal gluon propagator.  After explicitly
identifying the antiquark contribution, i.e.
 $W_{\ov{q}q}(k_-)=-W_{q\ov{q}}^T(-k_-)$, we perform the temporal
integration over the quark and antiquark propagators, which now leads
to (unlike \eq{eq:tempint}):
\bea
\frac{\imath}{2\pi}\int_{-\infty}^\infty
\frac{dk_0}{\left[k_+^0-m-{\cal I}_r+\imath\e\right]\left[k_-^0-m+
{\cal I}_r-\imath\e\right]}=\frac{-1}{{P_0-2\cal I}_r+2\imath\e}.
\label{eq:nonvanishing}
\eea

Further, we insert the expression \eq{eq:iregularized} for ${\cal
I}_r$ and after Fourier transforming to coordinate space we find the
following solution for the bound state energy of the quark-antiquark
system:
\be
P_0=g^2\int_{r}\frac{\dk{ \vec{\w}}D_{\si\si}
(\vec{\w})}{\vec{\w}^2}
\left[C_F-e^{\imath\vec{\w}\cdot\vec{x}}C_M\right]+{\cal O}
\left(1/m\right).
\label{eq:p0sol}
\ee
In the above, $C_M$ is an (unknown) color factor assigned to the \BS
vertex $\G$, which has to be yet identified:
\be
\left[T^a\G(\vec{x})T^a\right]_{\al\ba}=C_M\G_{\al\ba}(\vec{x}).
\ee 

Because the total color charge of the system is conserved and
vanishing \cite{Reinhardt:2008pr}, a quark cannot exist as an
asymptotic state.  Hence, the bound state energy $P_0$ of a
quark-antiquark system can be either infinite, such that the system is
not allowed, or linearly rising, if the system is confined. If the
temporal gluon propagator is more infrared divergent than
$1/|\vec{\w}|$, we find that in order to ensure the convergence of the
spatial integral, $C_M$ must be equal to $C_F$. This immediately leads
to the condition
\be
\G_{\al\ga}(\vec{x})=\de_{\al\ga}\G(\vec{x}),
\ee
which implies that the \BS equation can only have a finite solution
for \emph{color singlet} states. Further, if we assume that in the
infrared $D_{\si\si}=X/\vec{\w}^2$ (as indicated by the lattice
\cite{Quandt:2008zj}), where $X$ is a combination of constants, then
from \eq{eq:p0sol} we find
\be
P_0\equiv\si|\vec{x}|=\frac{g^2C_FX}{8\pi}|\vec{x}|+{\cal O}
\left(1/m\right).
\ee
This result shows that there is a direct connection between the
physical string tension $\si$ and the nonperturbative \YM sector of
QCD, at least under the truncation scheme employed here, which
corresponds to a color singlet bound state of a quark and an
antiquark, and otherwise the system has infinite energy.

\section {Four-point quark-antiquark Green's functions}
\label{sec:4pgf}

\begin{figure}[t]
\includegraphics[width=1.0\linewidth]{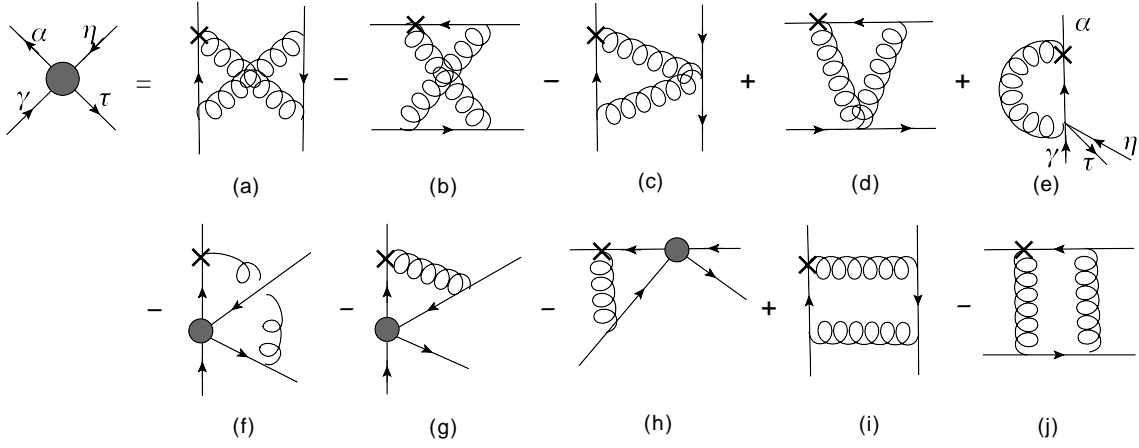}
\caption{\label{fig:1PI}
Diagrammatic representation of the \DS equation for the 1PI 4-point
quark-antiquark Green's function. Blobs represent dressed proper (1PI)
4-point vertex, solid lines represent the quark propagator, springs
denote either spatial ($\vec A$) or temporal ($\si$) gluon propagator
and cross denotes the tree level quark-gluon vertex. Internal
propagators and 1PI vertices are fully dressed.}
\end{figure}

Similar to the gap equation, we derive the the full \DS equation for
the proper (1PI) four-point quark-antiquark Green's function in
configuration space (see Ref.~\cite{arXiv:1103.4786} for notation and
technical details of the functional derivation):
\bea
&&\ev{\imath\ov{q}_{\al x}\imath q_{\ga z}\imath\ov{q}_{\tau w}
\imath q_{\eta t}}=
[g\ga^0T^a]_{\al\ba}\int dy\, \de(x-y)\times \nonumber\\
&&\left\{ 
\left[ \ev{\imath\ov\chi_{\ba x}\imath\chi_{\ka}}
\ev{\imath\ov{q}_{\ka}\imath q_{\ga z}\imath \si^{c}_{\la}}
\ev{\imath\rho^c_{\la}\imath\rho^d_{\de}} \right]\! \left[
\ev{\imath\ov{q}_{\tau w}\imath q_{\nu}\imath \si^{b}_{\mu}}
\ev{\imath\ov\chi_{\nu}\imath\chi_{\e}} \ev{\imath\ov{q}_{\e}\imath
q_{\eta t}\imath \si^{d}_{\de}} \ev{\imath\rho^b_{\mu}\imath\rho^a_y}
\right]\right.\nonumber\\
&&-
\left[ \ev{\imath\ov\chi_{\ba x}\imath\chi_{\de}}
\ev{\imath\ov{q}_{\de}\imath q_{\eta t}\imath \si^{c}_{\e}}
\ev{\imath\rho^c_{\e}\imath\rho^d_{\ka}} \right]\! \left[
\ev{\imath\ov{q}_{\tau w}\imath q_{\nu}\imath \si^{b}_{\mu}}
\ev{\imath\ov\chi_{\nu}\imath\chi_{\la}} \ev{\imath\ov{q}_{\la}\imath
q_{\gamma z}\imath \si^{d}_{\ka}}
\ev{\imath\rho^b_{\mu}\imath\rho^a_y} \right]\nonumber\\
&&-
\left[ \ev{\imath\ov\chi_{\ba x}\imath\chi_{\ka}}
\ev{\imath\ov{q}_{\ka}\imath q_{\ga z}\imath \si^{c}_{\la}}
\ev{\imath\rho^c_{\la}\imath\rho^d_{\nu}} \ev{\ov{q}_{\tau w}\imath
q_{\eta t}\imath \si^{d}_{\nu}\imath \si^{b}_{\mu}}
\ev{\imath\rho^b_{\mu}\imath\rho^a_y} \right]\nonumber\\
&&+
\left[ \ev{\imath\ov\chi_{\ba x}\imath\chi_{\de}}
\ev{\imath\ov{q}_{\de}\imath q_{\eta t}\imath \si^{c}_{\e}}
\ev{\imath\rho^c_{\e}\imath\rho^d_{\ka}}\right]\!
\left[\ev{\imath\ov{q}_{\tau w}\imath q_{\ga z}\imath\si^d_{\ka}
\imath\si^b_{\la}} \ev{\imath\rho^b_{\la}\imath\rho^a_y}
\right]\nonumber\\
&&+
\left[ \ev{\imath\ov\chi_{\ba x}\imath\chi_{\ka}}
\ev{\imath\ov{q}_{\ka}\imath q_{\ga z}\imath\ov{q}_{\tau w}\imath
q_{\eta t}\si^{b}_{\la}} \ev{\imath\rho^b_{\la}\imath\rho^a_y}
\right]\nonumber\\
&&-
\left[ \ev{\imath\ov\chi_{\ba x}\imath\chi_{\ka}}
\ev{\imath\ov{q}_{\ka}\imath q_{\ga z}\imath\ov{q}_{\la}\imath q_{\eta
t}} \ev{\imath\ov{q}_{\tau w}\imath q_{\nu}\imath \si^{b}_{\mu}}
\ev{\imath\ov\chi_{\nu}\imath\chi_{\la}}
\ev{\imath\rho^b_{\mu}\imath\rho^a_y} \right]\nonumber\\
&&-
\left[ \ev{\imath\ov\chi_{\ba x}\imath\chi_{\ka}}
\ev{\imath\ov{q}_{\ka}\imath q_{\ga z}\imath\ov{q}_{\tau w}\imath
q_{\la}} \ev{\imath\ov\chi_{\la}\imath\chi_{\de}}
\ev{\imath\ov{q}_{\de}\imath q_{\eta t}\imath \si^{b}_{\e}}
\ev{\imath\rho^b_{\e}\imath\rho^a_y} \right]\nonumber\\
&&-
\left[ \ev{\imath\ov\chi_{\ba x}\imath\chi_{\nu}}
\ev{\imath\ov{q}_{\nu}\imath q_{\mu}\imath\ov{q}_{\tau w}\imath
q_{\eta t}} \ev{\imath\ov\chi_{\mu}\imath\chi_{\ka}}
\ev{\imath\ov{q}_{\ka}\imath q_{\ga z}\imath \si^{b}_{\la}}
\ev{\imath\rho^b_{\la}\imath\rho^a_y} \right]\nonumber\\
&&+\left[
\ev{\imath\ov\chi_{\ba x}\imath\chi_{\ka}}
\ev{\imath\ov{q}_{\ka}\imath q_{\ga z}\imath \si^{c}_{\la}}
\ev{\imath\rho^c_{\la}\imath\rho^d_{\nu}} \right] \!\left[
\ev{\imath\ov{q}_{\tau w}\imath q_{\mu}\imath \si^{d}_{\nu}}
\ev{\imath\ov\chi_{\mu}\imath\chi_{\de}} \ev{\imath\ov{q}_{\de}\imath
q_{\eta t}\imath \si^{b}_{\e}} \ev{\imath\rho^b_{\e}\imath\rho^a_y}
\right]\nonumber\\
&&-
\left.  \left[ \ev{\imath\ov\chi_{\ba x}\imath\chi_{\de}}
\ev{\imath\ov{q}_{\de}\imath q_{\eta t}\imath \si^{c}_{\e}}
\ev{\imath\rho^c_{\e}\imath\rho^d_{\nu}} \right]\!\! \left[
\ev{\imath\ov{q}_{\tau w}\imath q_{\mu}\imath \si^{d}_{\nu}}
\ev{\imath\ov\chi_{\mu}\imath\chi_{\ka}} \ev{\imath\ov{q}_{\ka}\imath
q_{\ga z}\imath \si^{b}_{\la}} \ev{\imath\rho^b_{\la}\imath\rho^a_y}
\right] \right\} \nonumber\\
&&+\dots
\label{eq:1pi_4quark}
\eea
where the dots represent the $\vec A$ vertex terms (which will be
truncated out in our scheme), and we have already replaced the
tree-level temporal quark-gluon vertex with its expression
\eq{eq:feyn}.  The above equation is diagrammatically represented in
\fig{fig:1PI}.

We now proceed by applying our truncation scheme at leading order in
the mass expansion.  Since we shall consider the flavor non-singlet
Green's function in the $s$-channel (the quark and the antiquark are
regarded as two distinct flavors, but with equal masses), the diagrams
(a), (c) and (i) of \fig{fig:1PI} are excluded.  In the diagram (b)
(crossed ladder type exchange diagram), we insert as before the
appropriate propagators Eqs. (\ref{eq:quarkpropnonpert},
\ref{eq:antiquarkpropnonpert}) and vertices Eqs. (\ref{eq:feyn},
\ref{eq:antiqqsinp}).  The resulting energy integral is similar to the
integral (2.14) and vanishes, just like the higher order contributions
to the kernel of the homogeneous \BS equation.  Turning to the diagram
(d), we see that this contribution involves a quark-two gluon
vertex. From the corresponding \ST identity (derived explicitly in
Ref.~\cite{arXiv:1103.4786}), it can be seen that this vertex is zero
and hence the diagram (d) vanishes.

\begin{figure}[t]
\vspace{0.5cm}
\centering\includegraphics[width=0.8\linewidth]{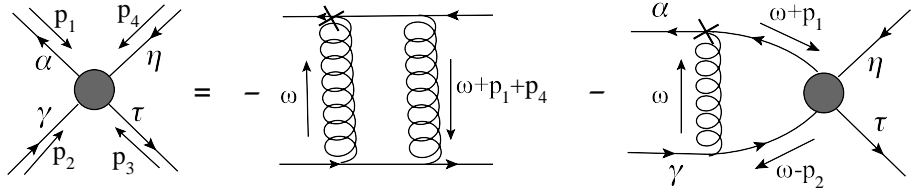}
\caption{\label{fig:1PI_mr}
Truncated \DS equation for the 1PI 4-point Green's function in the
$s$-channel. Same conventions as in  Fig.~2 apply.}
\end{figure}

At this stage, we adopt the following strategy: discard for the moment
the diagrams (f) and (g), which include the 1PI four-point
quark-antiquark Green's function, and the diagram (e), containing a
four-quark-gluon vertex, solve the equation with the remaining terms
(the diagram (h) and the rainbow-ladder term (j)), and with the
obtained solution return to the diagrams (f), (g) and (e), and show
that they cancel (and hence our assumption is consistent).  In this
case, \eq{eq:1pi_4quark} reduces to the \DS equation for the 1PI
four-point quark Green's function in the $s$-channel, in the
ladder-approximation (shown diagrammatically in \fig{fig:1PI_mr}):
\bea
\lefteqn{ \G^{(4)}_{\al\ga\tau\eta}(p_1,p_2,p_3,p_4)=
}\nonumber\\
&&-\int\dk{\w}\left[\G_{\bar qq\si \al\ba}^{(0)a}(p_1,-p_1-\w,\w)
W_{\bar qq \ba\de}(p_1+\w) \G_{\bar qq\si
\de\eta}^{c}(p_1+\w,p_4,-p_1-p_4-\w)\right]
\nonumber\\
&&\times\left[\G_{\bar qq\si\tau\mu}^{d}(p_3,p_2-\w,p_1+p_4+\w)
W_{\bar qq\mu\ka}(\w-p_2) \G_{\bar
qq\si\ka\ga}^{b}(\w-p_2,p_2,-\w)\right]\nonumber\\ 
&& \times W_{\si\si}^{ab}(- \w)
W_{\si\si}^{cd}(p_1+ p_4+\w)
\nonumber\\
&&-\int\dk{\w}\G_{\bar qq\si\al\ba}^{(0)a} (p_1,-p_1-\w,\w) W_{\bar qq
\ba\nu}(p_1+\w) \G^{(4)}_{\nu\mu\tau\eta}(p_1+\w,p_2-\w,p_3,p_4)
W_{\bar qq\mu\ka}(\w-p_2)\nonumber\\
&&\times \G_{\bar qq\si \ka\ga}^{b} (\w-p_2,p_2,-\w)
W_{\si\si}^{ab}(-\w).
\label{eq:4pointDS1}
\eea

As before, we identify the antiquark component of \eq{eq:4pointDS1}
(lower line of \fig{fig:1PI_mr}) and insert the expressions
Eqs.~(\ref{eq:quarkpropnonpert}, \ref{eq:antiquarkpropnonpert}), for
the quark and antiquark propagators, along with the vertices
Eqs.~(\ref{eq:feyn}, \ref{eq:antiqqsinp}) and the definition
\eq{eq:Wsisi} for the temporal gluon propagator.  We further make the
assumption that $\G ^{(4)}(p_1,p_2,p_3,p_4)=\G ^{(4)}(P_0;\vec
p_1+\vec p_4)$, where $P_0=p_1^0+p_2^0$, which allows us to separate
the three-momentum and energy integrals.  The energy integrals are
similar to \eq{eq:nonvanishing} and can be carried out.  Making the
following color decomposition for the function $\G^{(4)}$:
\be
\G^{(4)}_{\al\ga\tau\eta}
=\de_{\al\ga}\de_{\tau\eta}\G_{1}^{(4)}+\de_{\al\eta}\de_{\tau
\ga}\G_{2}^{(4)},
\label{eq:g4color1PI}
\ee
where $\G^{(4)}_{1}$ and $\G^{(4)}_{2}$ are scalar functions and
Fourier transforming back to coordinate space, we find the following
solution for the 1PI quark-antiquark Green's function:
\bea
\G^{(4)}_{\al\ga\tau\eta} (P_0; x)
&=&\imath\left(\frac{g^2}{2N_c}\right)^2 \frac{W_{\si\si}(x)^2}
{P_0-2{\cal I}_r+\imath \frac{g^2}{2N_c}W_{\si\si}(x)
+2\imath\e}\nonumber\\
&&\times\left\{ \de_{\al\ga}\de_{\tau\eta} \frac{(P_0-2{\cal
I}_r)N_c(N_c^2-2)+\imath g^2N_c C_F W_{\si\si}(x)} {P_0-2{\cal
I}_r-\imath g^2C_FW_{\si\si}(x) +2\imath\e}
+\de_{\al\eta}\de_{\tau\ga} \right\},
\label{eq:sol1PI}
\eea
where $x=|\vec x|$ is the separation associated with the momentum
$\vec p_1+\vec p_4$.

As promised, with the solution \eq{eq:sol1PI} for the 1PI Green's
function, we now return to the diagrams (f), (g) and (e) and show that
they do not contribute to the final result. To see this, we first
consider the diagram (g) and notice that the energy dependence of the
internal four-point function can be written as
\be
\G^{(4)}(P_0+\w_0)\sim \frac{\w_0^m}{\left[\w_0
+X+\imath \e\right]^n},
\ee
where $X$ is a combination of constants, $n=1,2$ and $m=0,1$. Then the
energy integral takes the form
\be
\int\dk{\w_0}\w_0^m\prod_{i=1}^{2+n}\frac{1}{\left[\w_0
+X_i+\imath\e\right]}=0.
\label{eq:loopcorr}
\ee
Clearly, this integral is a generalization of \eq{eq:tempint} and this
vanishes, just as for the loop corrections in the kernel of the \BS
equation from the previous section and the diagram (b) from above.  An
identical calculation for the diagram (f), recalling that the lower
line corresponds to an antiquark propagator, leads us to the fact that
this integral is also zero.  Finally, turning to the diagram (e),
containing the four quark-gluon vertex, we notice that the
perturbative series of this diagram coincides with the ladder
resummation of the diagrams (f) and (g), which we have found to be
vanishing, and hence this diagram is also zero (even though the
five-point interaction vertex itself does not vanish -- see also
Ref.~\cite{arXiv:1103.4786} for a detailed diagrammatic analysis).  In
turn, this implies that our original assumption is correct and the
solution \eq{eq:sol1PI} is valid at every order in perturbation
theory. Marginally, we note that the fact that the five-point function
is finite relates to the existence of three-quark bound states in the
Faddeev equation, in the ladder approximation \cite{arXiv:1010.4254}.

\begin{figure}[t]
\centering\includegraphics[width=0.8\linewidth]{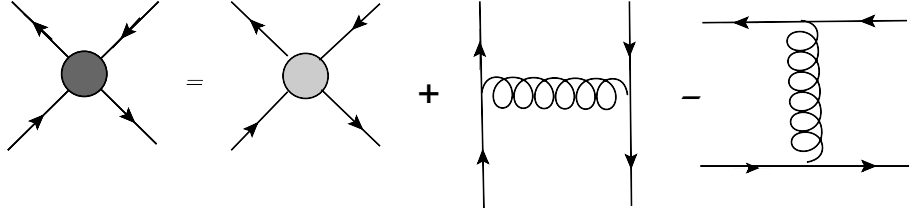}
\caption{\label{fig:1PI_amputated}
Relation between the 1PI (dark blob) and amputated (shaded blob)
4-point Greens function for the quark-antiquark system. Internal
propagators and 1PI vertices are fully dressed.}
\end{figure}

Let us now consider the \DS equation for the fully amputated
four-point quark-antiquark Green's function in the $s$-channel, which
we denote $G^{(4)}$. This is related to the 1PI function $\G^{(4)}$
via the Legendre transform (see \fig{fig:1PI_amputated}).  This study
is motivated by the fact that this equation reduces (under truncation)
to the inhomogeneous ladder \BS equation, from which the homogeneous
\BS equation of the previous section is derived. In the following, we
will derive the solutions of this equation, analyze the positions of
the poles, and explicitly verify that the physical solutions coincide
with the bound state solutions of the homogeneous \BS equation.

\begin{figure}[t]
\centering\includegraphics[width=0.75\linewidth]{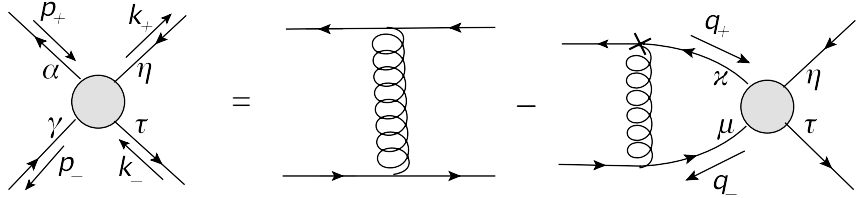}
\caption{\label{fig:amputated_mr}
Truncated \DS equation for the fully amputated
quark-antiquark 4-point Green's function in the $s$-channel.}
\end{figure}

Starting with \eq{eq:4pointDS1} (in coordinate space) for the proper
function, we replace the 1PI function $\G^{(4)}$ with the amputated
function $G^{(4)}$ according to \fig{fig:1PI_amputated}, and cut the
external quark legs.  The resulting equation for $G^{(4)}$ reads (see
also \fig{fig:amputated_mr}):
\bea
&& G^{(4)}_{\al\ga;\tau\eta}(p_+,p_-;k_+,k_-)= W_{\si\si}^{ab}(\vec
p-\vec k)\left[\G_{\bar qq\si}^{a}\right]_{\al\eta} \left[\G_{\bar q
q\si}^{b}\right]_{\tau\ga}
\nonumber\\
&&-\int\dk{q}\left[\G_{\bar qq\si}^{a(0)} W_{\bar
qq}(q_+)\right]_{\al\ka} \left[W_{\bar qq}(q_-)\G_{\bar
qq\si}^{b}\right]_{\mu \ga} W_{\si\si}^{ab}(\vec p-\vec q)
G^{(4)}_{\ka\mu;\tau\eta}(q_+,q_-;k_+,k_-).
\label{eq:4pointDS2}
\eea
In the above, for the quark momenta we use the same conventions as in
the \BS equation, \eq{eq:bseq}.  Again, we replace the heavy quark and
antiquark propagators and vertices with the expressions
Eqs.~(\ref{eq:quarkpropnonpert}, \ref{eq:antiquarkpropnonpert},
\ref{eq:feyn}, \ref{eq:antiqqsinp}), perform the energy integration,
and Fourier transform back to coordinate space.  Similar to the proper
four-point function, we make a color decomposition of the function
$G^{(4)}$:
\be
G^{(4)}_{\al\ga;\tau\eta} =\de_{\al\ga}\de_{\tau
\eta}G^{(4)}_{1}+\de_{\al\eta}\de_{\tau \ga}G^{(4)}_{2},
\label{eq:g4color}
\ee
($G^{(4)}_{1}$ and $G^{(4)}_{2}$ are scalar functions), and after
using the Fierz identity to sort out the color factors, we obtain the
final result for the function $G^{(4)}$:
\bea
&G^{(4)}_{\al\ga;\tau\eta}(P_0;x)&=\frac{g^2}{2} \frac{\left(P_0-2{\cal
I}_r\right) W_{\si\si}(x)}{P_0-2{\cal I}_r+\imath
\frac{g^2}{2N_c}W_{\si\si}(x) +2\imath\e} \nonumber\\
&&\times\left[
\de_{\al\ga}\de_{\tau\eta} \frac{(P_0-2{\cal I}_r) } {P_0-2{\cal
I}_r-\imath g^2C_F W_{\si\si}(x) +2\imath\e}-
\de_{\al\eta}\de_{\tau\ga}\frac{1}{N_c} \right],
\label{eq:4pointfinal}
\eea
where  $P_0=p_{+}-p_{-}$ is the total energy of the $\bar qq$ state.

Analyzing the structure of the four-point functions, we first notice
that even though the results, Eqs.~(\ref{eq:sol1PI},
\ref{eq:4pointfinal}), are derived under truncation, the denominator
structure of the 1PI and amputated Green's functions are identical in
both color channels, and furthermore, the physical and nonphysical
singularities have disentangled automatically.  Using the form
\eq{eq:Wsisi} for the temporal gluon propagator, the denominator
factor of the color singlet channel in either \eq{eq:sol1PI} or
\eq{eq:4pointfinal} can be rewritten in the form
\be
P_0-g^2\int_{r}\frac{\dk{\vec{\w}}D_{\si\si}(\vec{\w})}{\vec{\w}^2}
C_F\left[1-e^{\imath\vec{\w}\cdot\vec{x}}\right].
\ee
In this expression we recognize the bound state (infrared confining)
energy $P_{0\; res}(x)=\si|\vec x|$ (further assuming that
$D_{\si\si}(\vec\w^2)\sim 1/\vec\w^2$), similar to \eq{eq:p0sol} for
the homogeneous \BS equation in the color-singlet channel.  Hence, we
have found an explicit analytical dependence of the four-point Green's
function on the $\bar qq$ bound state energy resulting from the
homogeneous \BS equation.

Turning to the overall denominator factors of Eqs.~(\ref{eq:sol1PI},
\ref{eq:4pointfinal}), i.e. the denominator factor not specific to the
color-singlet channel, we again insert the explicit form of the
temporal gluon propagator and arrive at the following result:
\be
P_0-g^2\int_{r}\frac{\dk{\vec{\w}}D_{\si\si}(\vec{\w})}{\vec{\w}^2}
\left[C_F+\frac{1}{2N_c}e^{\imath\vec{\w}\cdot\vec{x}}\right].
\ee
This factor does not appear in the homogeneous \BS equation; it is
part of the normalization and, similar to the quark propagator,
represents an unphysical pole which is shifted to infinity when the
(implicit) regularization is removed.

\section{Conclusions}

In this talk, we have discussed the \DS and \BS equations for
quark-antiquark systems in Coulomb gauge, at leading order in the
heavy quark mass expansion and with the truncation to include only the
(nonperturbative) temporal gluon propagator.  Under this truncation,
the rainbow approximation to the quark gap equation is exact, as is
the corresponding ladder approximation to the homogeneous \BS
equation.  The only physical solution corresponds to confinement,
i.e. only color singlet meson states have finite energy (and hence are
physically allowed), and otherwise the system has infinite
energy. Incidentally, these results are supported by recent \DS
studies in Coulomb gauge at leading order \cite{Watson:2011kv}.

Turning to the four-point quark-antiquark Green's functions, we have
presented analytic solutions for both the proper and amputated Green's
functions. The two functions have the same denominator structures, and
the physical and nonphysical singularities disentangle, the physical
poles coinciding with the bound state solutions obtained for the
homogeneous \BS equation.

\acknowledgments
C.P. has been supported by the Deutscher Akademischer Austausch Dienst
(DAAD), the Helm{\-}holtz International Center for FAIR within the
LOEWE program of the State of Hesse, and the Helmholtz Young
Investigator Group No.~VH-NG-332.  P.W. and H.R. have been supported
by the Deutsche Forschungsgemeinschaft (DFG) under contracts
no. DFG-Re856/6-2,3. C.P.  thanks the organizers, in particular
C. Aguilar and D. Binosi, for the support.


\end{document}